\documentclass[12pt]{iopart}

\usepackage{graphics}
\usepackage{subfig}
\usepackage{color}
\begin{document}

\title[Static properties of cluster formations in a multiblock copolymer chain]{Analysis of the static
properties of cluster formations in symmetric linear multiblock copolymers}

\author{N G Fytas$^1$ and P E Theodorakis$^2$$^,$$^3$$^,$$^4$}

\address{$^1$ Department of Materials Science, University of Patras, 26504 Patras, Greece}
\address{$^2$ Faculty of Physics, University of Vienna, Boltzmanngasse 5, A-1090 Vienna, Austria}
\address{$^3$ Institute for Theoretical Physics and Center for Computational Materials Science, Vienna University of Technology, Hauptstra{\ss}e 8-10, A-1040 Vienna, Austria}
\address{$^4$ Vienna Computational Materials Laboratory, Sensengasse 8/12, A-1090 Vienna, Austria}

\ead{nfytas@phys.uoa.gr}
\ead{panagiotis.theodorakis@univie.ac.at}

\begin{abstract}
We use molecular dynamics simulations to study the static
properties of a single linear multiblock copolymer chain under
poor solvent conditions varying the block length $N$, the number
of blocks $n$, and the solvent quality by variation of the
temperature $T$. We study the most symmetrical case, where the
number of blocks of monomers of type A, $n_{A}$, equals that of
monomers B, $n_{B}$ ($n_{A}=n_{B}=n/2$), the length of all blocks
is the same irrespective of their type, and potential parameters
are also chosen symmetrically, as for a standard
Lennard-Jones fluid. Under poor solvent conditions the chains
collapse and blocks with monomers of the same type form clusters,
which are phase separated from the clusters with monomers of the
other type. We study the dependence of the size of the formed
clusters on $n$, $N$ and $T$. Furthermore, we discuss our results with respect
to recent simulation data on the phase behaviour of such
macromolecules, providing a complete picture for the cluster
formations in single multiblock copolymer chains under poor
solvent conditions.
\end{abstract}


\submitto{\JPCM}

\maketitle

\section{Introduction}
\label{intro}

Block copolymers have been recently the subject of many
experimental and theoretical studies, as such systems are highly
involved in industrial
applications~\cite{1,2,3,4,5,6,7,8,9,10,11,12,13,14,15,16,17,18,19,20,21,22,23,24,25,26,27,28,29,30,31,32,33,
34,35,36,37,38,39,40,41,42,43,44,45,46,47,48,49,50,51,52,53,54,55,56}.
Block copolymer melts is the most studied system with this
respect~\cite{1,2,3,4,5,6,7,8,9,10,11,12,13,14,15,16,17,18,19,20,21,22,23,24,25,26,27,28,29,30,31,32,33,
34,35,36,37,38,39,40,41,42,43,44,45,46,47,48,49}. We also note
here that various copolymer systems have been studied in many
other occasions concerning different geometries, i.e., in a
selective nanoslit~\cite{57}, on nanopatterned
surfaces~\cite{58,59,60}, in nanopores~\cite{61}, in the case of
nanoparticle aggregation~\cite{62}, etc. Lots of theoretical,
experimental, and simulation studies are dealing with these
systems. In particular, theory predicts that melts of block
copolymer chains for the most symmetrical case (equal block length
and composition) form lamellar structure, given that the scaling
parameter that controls the phase behaviour ($\chi N$) is high
enough to lead to phase separation between the A and B
blocks~\cite{11,12,13}. Simulations have confirmed that this
parameter controls the phase behaviour of such systems~\cite{4}.
It has also been discussed that the geometry of the microphase
separated regions is controlled by the number of blocks $n$, as
well as other parameters, i.e., relative size and arrangement of
blocks~\cite{4}. Then, theoretical studies were extended covering
the case of melts of multiblock copolymer chains (two types of
monomers are composing each chain composed of more than two
blocks). It was shown that this case has similar behaviour to a
melt of diblock copolymer chains~\cite{36}. That is, a lamellar
structure for the most symmetric case. Nevertheless, in the case
of infinitely dilute solutions, it is sufficient to study isolated
multiblock copolymer chains, where interactions (which can be of
short range) only within the chain and the effect of the solvent
are relevant~\cite{50,51,52,53,54,55,56}. The phase behaviour of
such systems has been studied by means of computer simulations for
chain lengths and temperature ranges accessible to
simulations~\cite{50,51}. Also, the dynamic properties of such
macromolecules have been the subject of recent simulation
studies~\cite{55}. Interestingly, such systems are also closely
related to various toy models (i.e., the HP model~\cite{63}),
which try to mimic the behaviour of various biomacromolecules,
which are formed by periodically repeated structural units
(``monomers'') along their chain, in order to understand
complicated biological processes, i.e., protein folding~\cite{64},
helical structures~\cite{65} etc.

Under Theta conditions ($\Theta$) or temperatures close to the
$\Theta$ temperature, it is expected that a single linear
symmetric multiblock copolymer chain would form a coil structure,
where blocks of different type (A,B) hardly come
across~\cite{50,51,52,53,54,55,56}. In fact, an expansion in the
chain dimensions of the chain is taking place due to the repulsive
interactions between neighboring blocks A and B along the chain.
Also, the spherical symmetry of the macromolecule should give its
place to an ellipsoidal overall formation~\cite{52,53,54,55,56}.
The chain flexibility could also be restricted as a result of the
neighboring unfavorable interactions of different monomer types
along the chain. However, the most interesting behaviour is
observed under poor solvent conditions. In this case,
chain collapses adopting a globular overall formation and different aspects of phase behaviour
have been discussed depending on the block length $N$, the number
of blocks $n$, and the temperature $T$, which can be used to tune
the quality of the solvent~\cite{50,51}. It has been discussed
that three different regimes can be unambiguously distinguished; a
regime where only two domains of monomers of different type are
formed. In this case all blocks of A monomers form a single
cluster, while all the blocks of B-type monomers belong to another
cluster without occurrence of any variation in the number of
clusters. This means that the clusters are always separated by an
A-B interface and never one cluster of A or B monomers splits in
two distinct clusters. Another scenario suggests that full phase
separation, as defined above, takes place with a certain
probability, which can be very high or very small according to the
values of $N$, $n$, and $T$, while the third scenario corresponds to the
case that full phase separation as discussed
above (formation of only two clusters with different type of
monomers) can not take place and a symmetric variation in the number of clusters
$N_{cl}$ around an average value $(2<N_{cl}<n)$ is
observed~\cite{50,51}.

Guided by
theory~\cite{36}, one would rather expect in the long chain limit
that a ground-state type structure would be a single lamellar
domain, where an interface between all A- and B-type blocks is
formed, similarly to what is known for multiblock copolymer melts.
Such a structure would have much less (unfavorable) A-B contacts
compared to a multidomain structure of A and B clusters, which is
kinetically favored in simulations. The phase behaviour of
multiblock copolymer chains has been already rather extensively
discussed~\cite{50,51}. The overall properties of single
multiblock copolymer chains have also been the subject of various
studies~\cite{52,53,54,55,56}. In the following we focus our
discussion on the static properties of the formed clusters of a
single multiblock copolymer chain for a variety of parameters $N$,
$n$, and $T$ accessible to our simulations. In this way, we
provide a complete picture of the cluster formations in a single
multiblock copolymer chain under poor solvent conditions. Our
results are discussed within the framework of recent results on
the phase behaviour of such macromolecules~\cite{50,51}.

The rest of the paper is organized as follows: In Sec.~\ref{model}
we describe our simulation model and methods to analyze our
results. In Sec.~\ref{results} we present our results on the
analysis of the size of the formed clusters, and in
Sec.~\ref{conclusions} we give our concluding remarks.

\section{Model and methods to analyze the results}
\label{model}

\begin{figure*}
\begin{center}
\rotatebox{0}{\resizebox{!}{0.15\columnwidth}{%
  \includegraphics{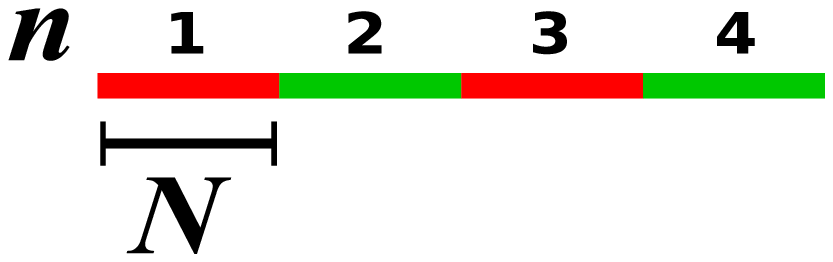}
}}
\end{center}
\caption{\label{fig1}(Colour online) Definition of structural parameters describing our linear
multiblock copolymer chains. $n$ (in this case $n=4$) is the number of different blocks
A and B ($n_{A}=n_{B}=n/2$) denoted with different color (or grey tone)
and $N$ is the length of each block.
All the blocks, irrespective of whether they are of type A or B,
have the same length $N$. Then the total length of the chain is
$nN$.}
\end{figure*}

We have considered symmetric linear multiblock copolymers, i.e.,
the length of all blocks, irrespective of their type, is $N$, the
number of A-type blocks is equal to the number of B-type blocks
with $n$ being an even number denoting the total number of blocks.
In our model the blocks of type A and B alternate along the chain.
The above parameters are presented schematically in
figure~\ref{fig1}, where the different colours correspond to the
different types of blocks. A single multiblock chain of four
blocks is shown; two blocks of A-type monomers and two blocks of
B-type monomers compose the multiblock chain in this figure.

Our chains are modelled by the standard bead-spring
model~\cite{50,51,55,66,67,68,69,70,71}, where the beads interact
via a cut and shifted Lennard-Jones (LJ) potential given by the
following formula
\begin{equation}
\label{eq:1}
U_{LJ}^{\alpha \beta}(r)= 4 \varepsilon_{LJ}^{\alpha \beta} [(\sigma_{LJ}^{\alpha \beta}/r)^{12} -
(\sigma_{LJ}^{\alpha \beta}/r)^6] + C, \quad r \leq r_c \quad,
\end{equation}
where $\alpha$, $\beta=A, B$ denote the different type of
monomers, and the constant $C$ is defined such that the potential
is continuous at the cut-off ($r_{c}=2.5$). For simplicity,
$\sigma_{LJ}^{\alpha \beta}=1$, $k_{B}=1$, but
$\varepsilon_{LJ}^{AA}=\varepsilon_{LJ}^{BB}=2\varepsilon_{LJ}^{AB}=1$,
in order to create an unmixing tendency between monomers A and B
belonging to different blocks as done in previous studies, and as
is used for a standard system (LJ fluid) ~\cite{72}.
Therefore, $\Delta \varepsilon=
\varepsilon_{LJ}^{AB}-1/2(\varepsilon_{LJ}^{AA}+\varepsilon_{LJ}^{BB})$
was kept the same throughout our simulations and $\chi$ ($\chi
\propto \Delta \varepsilon /T$) was varied by tuning the
temperature $T$. The connectivity along the chain is maintained by
the ``finitely extensible non-linear elastic'' (FENE) potential
\begin{equation}
\label{eq:2}
U_{\rm FENE} =-\frac{1}{2} k r^2_0 \ln [1-(r/r_0)^2], \quad 0 < r
\leq r_0.
\end{equation}
$U_{FENE}(r\geq r_{0})= \infty$, and the standard choices
$r_0=1.5$ and $k=30$ were used.

\begin{figure}
\begin{center}
\rotatebox{0}{\resizebox{!}{0.90\columnwidth}{%
  \includegraphics{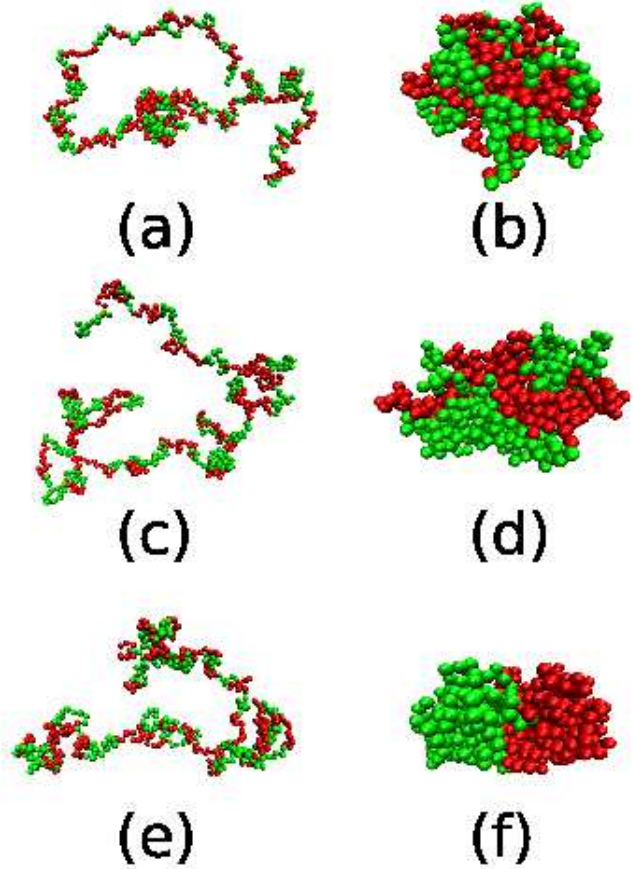}
}}
\end{center}
\caption{\label{fig2}(Colour online) Snapshot pictures of
different multiblock copolymers with the same total length
$nN=600$. (a) and (b) refer to a multiblock copolymer chain with
$N=6$, (c) and (d) to $N=15$, and (e) and (f) to $N=60$. Different
colours (or grey tone) correspond to different type of monomers.
Cases (a), (c), and (e) are snapshots taken at $T=3.0$. At this
temperature the snapshots look rather similar and the chain obtain
coil-like formations. Snapshots (b), (d), and (f) correspond to
temperature $T=1.5$. In this case the chain collapses and
different scenarios of phase behaviour are conceivable. In case (b) a
symmetric variation of the number of clusters ($N_{cl}$) around an
average value ($2<N_{cl}<n$) takes place during the simulation. The formation of only two
clusters containing different type of beads of all blocks never
occurs. In case (d) the formation of two only clusters (of
different beads) occurs with a certain probability, while in case
(f) this probability is $P(N_{cl}=n=2)$=1 (one cluster contains blocks of
with monomers of type A and one cluster is composed of blocks of B-type monomers).}
\end{figure}

For this model we know rather roughly the $\Theta$
temperature~\cite{73}, namely $\Theta=3.0$. We also know that for the
LJ fluid, which is a standard system, phase separation occurs at
a temperature close to $1.5$. Moreover, the phase
separation is favored by the increasing degree of polymerization, and
we expect to be able to observe phase separation at temperatures
down to $T=1.5$, making also use of our previous experience with
this model~\cite{74,75}. Indeed, phase
separation is encountered for the systems under consideration (e.g., see
figure~\ref{fig2}). For many of the studied cases phase
separation is already observed at temperatures close to $T=2.4$, which
is the temperature that the system enters to the regime where
the chains collapse, i.e., effects due to solvent are progressively
becoming important. It is
also clear that the use of symmetric structural and potential
parameters facilitates our study. Moreover comparison with
theoretical arguments could rather be more relevant. We use
standard molecular dynamics (MD) simulations where the temperature
is controlled by a Langevin thermostat, as is done in previous
studies~\cite{50,51,66,67,68,69,70,71,72,73,74,75}. Thus, the equation
of motion
\begin{equation}
\label{eq:3}
m \frac {d^2\vec{r}_i}{dt^2} = - \nabla U_i-m\gamma \frac
{d\vec{r}_i}{dt} + \vec{\Gamma}_i(t)
\end{equation}
is numerically integrated using the GROMACS package. In
Eq.~(\ref{eq:3}), $t$ denotes time, $U_i$ is the total potential
the $i$th bead experiences, $m$ is the mass which is taken as
unity, $\gamma$ is the friction coefficient,
and $\vec{\Gamma}_i(t)$ the random force. $\gamma$ and
$\vec{\Gamma}_i(t)$ are related by the standard fluctuation-dissipation
relation
\begin{equation}\label{eq:4}
\langle \vec{\Gamma}_i(t)\cdot \vec{\Gamma}_j(t')\rangle = 6 k_BT
\gamma \delta _{ij}\delta (t-t')\;.
\end{equation}
As in previous work~\cite{66,67,68,69,70,71,72,73,74,75}, the
friction coefficient was chosen as $\gamma = 0.5$. For the
integration of Eq.~(\ref{eq:3}) the leap-frog algorithm~\cite{76}
is used with a time step of $\Delta t = 0.006 \tau$, where the
natural time unit is defined as $\tau = (m \sigma
^2_{LJ}/\epsilon_{LJ})^{1/2}=1$.

We simulated our systems at a temperature close to the $\Theta$
temperature ($T \approx 3.0$) with an integration time step
$\Delta t=0.006 \tau$. For longer chains, we used  higher
temperatures for long equilibration runs, which were typically $30
\times 10^6 \tau$. After equilibration, we collected a number of
independent samples (typically $500$), which we used as initial
configurations for slow cooling runs. We remind the reader, that
in our simulations, the effect of the solvent is taken into
account only implicitly, i.e., the solvent quality is tuned by
variation of the temperature as is common practice. For our slow
cooling runs, the temperature was lowered from a high temperature
($T=3.0$) to a lower one ($T=1.5$) in temperature steps $\Delta
T=0.1$, for running the system at each temperature for a time
range of $2 \times 10^6$ MD steps. For temperatures below $T=2.1$
the intrinsic relaxation time of the chains starts to exceed the
simulation time. Therefore, taking this big number of
statistically independent ``cooling histories'' is indispensable,
in order to obtain meaningful statistical results. We also point
out that, the temperature $T=1.5$ is low enough in order to access
the most interesting regime where the chains collapse to form
cluster formations. Of course, simulating lower temperatures could
be also interesting in order to try to compare our results with
relevant theoretical arguments given our rather short chain
lengths, but then it would be impossible to simulate such a system
with our method. We also note, that the size $L$ of the simulation
box was chosen such that the multiblock copolymer chain never
interacts with its periodic images. For instance, for a chain of
total length $nN=600$, $L=120$. For smaller or longer chains we
used correspondingly smaller or bigger simulation boxes satisfying
always the above criterion for the size $L$ of the simulation box.
Further details on the simulated model have been discussed
elsewhere~\cite{50,51}.

At temperatures close to $\Theta$, coil structures are formed and
it seems visually (figure~\ref{fig2}(a), (c), and (e)) that
small differences exist for multiblock chains of
different structural parameters $n$, $N$, as for example in
figure~\ref{fig2}. Of course, even at this high temperature
differences in the overall formation of the coil structure occur,
but, here, we mainly focus on the static properties of the
clusters. The most interesting effect would be that the overall
formation of the chains deviates from a spherical shape. At such
temperatures, blocks of the same type of beads can only form
occasionally clusters due to the highly present thermal
fluctuations. At temperatures below $T \approx 2.4$ the chains collapse
and blocks of the same type of monomers form clusters with other
monomers belonging to different blocks, due to the unfavorable
interactions between A and B beads. Figure~\ref{fig2} shows also
this case of phase separation, where in case (b) all blocks of the
same type are not able to join together to a single cluster. In
this case, we have a symmetric variation in the number of
clusters, as it has been discussed elsewhere~\cite{50,51}. Then in
case (d), the formation of two only clusters of dissimilar blocks
is taking place with a certain probability, which might be high or
small according to the choice of parameters $n$ and $N$. In
case (f) the blocks of different type are always phase separated
and they form two different clusters, with an A-B interface formed
between clusters. Here, the thermal fluctuations cannot
overcome the incompatibility between dissimilar beads, and the
clusters are always phase separated. One would need to identify
the beads belonging to each cluster in order to describe the overall
size of these cluster formations. We used the Stillinger criterion
and the standard choice $r_n=1.5$~\cite{77}, as is done in
previous work~\cite{50,51,74,75}. We also used smaller values of
$r_n$ and similar results were obtained, while higher $r_n$ values
are hardly significant, due to the rapid fall-off of the LJ
potential.

Properties, $F$, depending on the fluctuating number of clusters,
should be averaged with the probability $P(x)$ that a number of
clusters per block occurs ($x$), i.e.,
\begin{equation}\label{eq:5}
\bar{F} = \sum \limits _{x}P(x) F(x).
\end{equation}
Due to the symmetry of our model,
the statistical analysis for A and B clusters should end
up in the same results. This is confirmed in our analysis,
validating also our simulation protocol. Therefore, results on the
clusters' properties will refer either to clusters with monomers
of type A or clusters of monomers B only, and nowhere clusters of
A and B monomers in a single cluster are considered, although this
is possible given our criterion $r_{n}$ due to thermal
fluctuations.

\section{Results and Discussion}
\label{results}

\begin{figure*}
\begin{center}
\subfloat[][]{
\rotatebox{270}{\resizebox{!}{0.50\columnwidth}{%
  \includegraphics{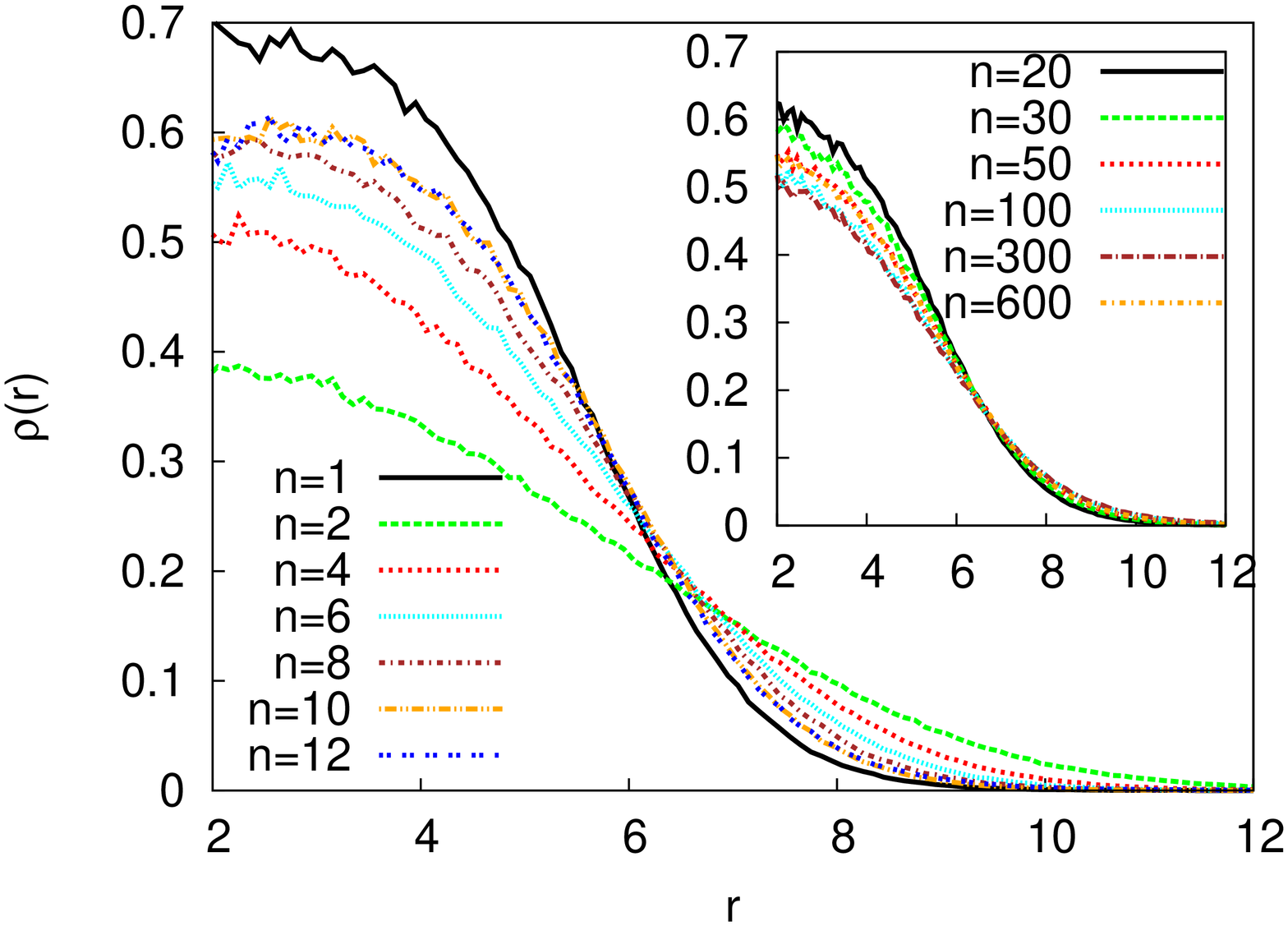}
}}}
\subfloat[][]{
\rotatebox{270}{\resizebox{!}{0.50\columnwidth}{%
  \includegraphics{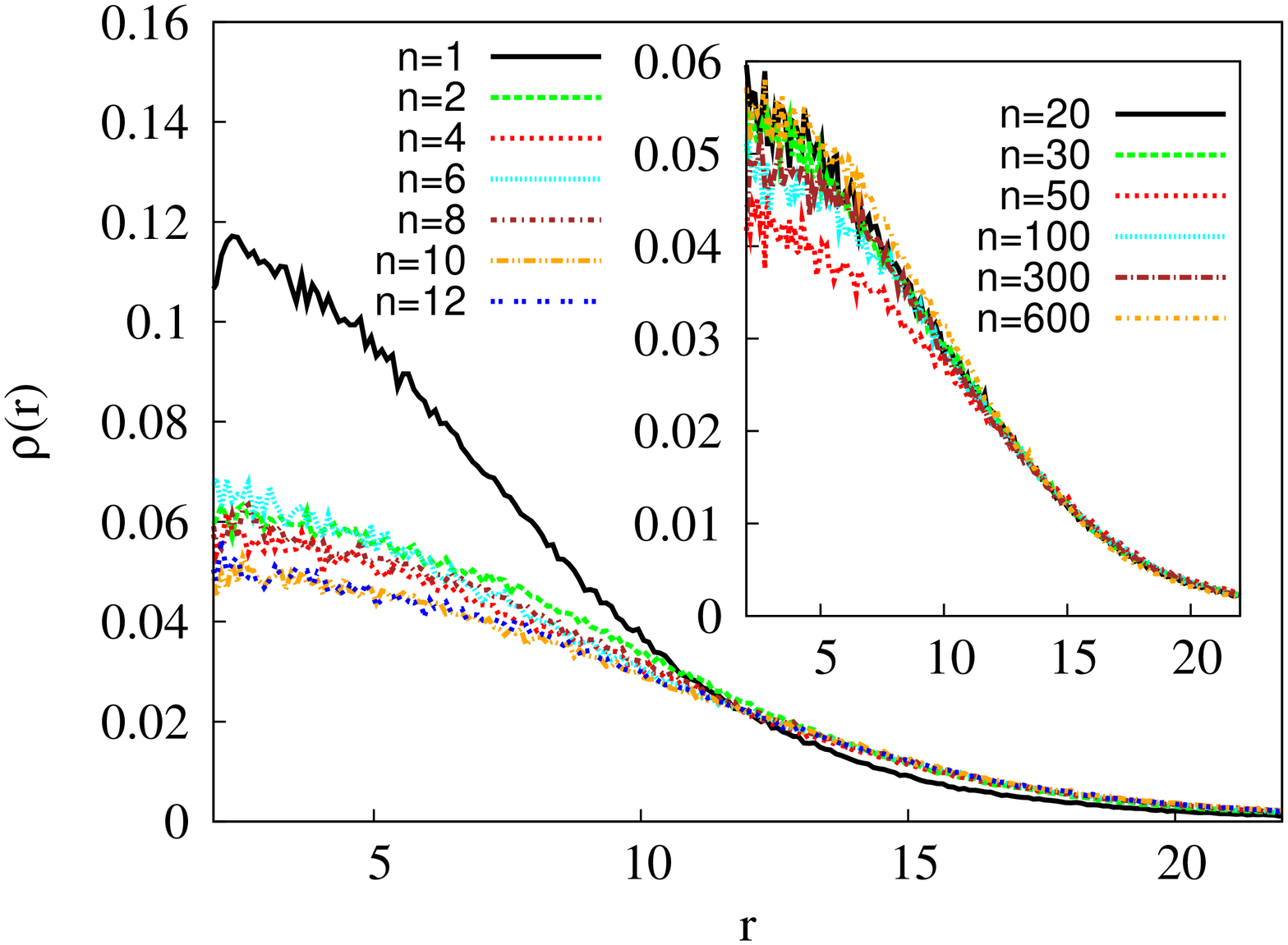}
}}}
\end{center}
\caption{\label{fig3}(Colour online) Density profile
$\rho(r)=\rho(|\vec{r}|)$ plotted versus the radial distance
$r=|\vec{r}|$ from the center of mass of the chain. The
temperatures $T=1.8$ (a) and $T=3.0$ (b) are shown.}
\end{figure*}

In figure~\ref{fig3} we present results for the density profiles
of multiblock chains with the same total length $nN=600$, but
different combination of parameters $n$ and $N$, mathematically
expressed by the following formula
\begin{equation}\label{eq:6}
\rho(|\vec{r}|)= \langle  \sum \limits_{i=1}^{nN} \delta(
\vec{r}-\vec{r}_{c}-\vec{r}_{i}) \rangle,
\end{equation}

\begin{figure*}
\begin{center}
\subfloat[][]{
\rotatebox{270}{\resizebox{!}{0.50\columnwidth}{%
  \includegraphics{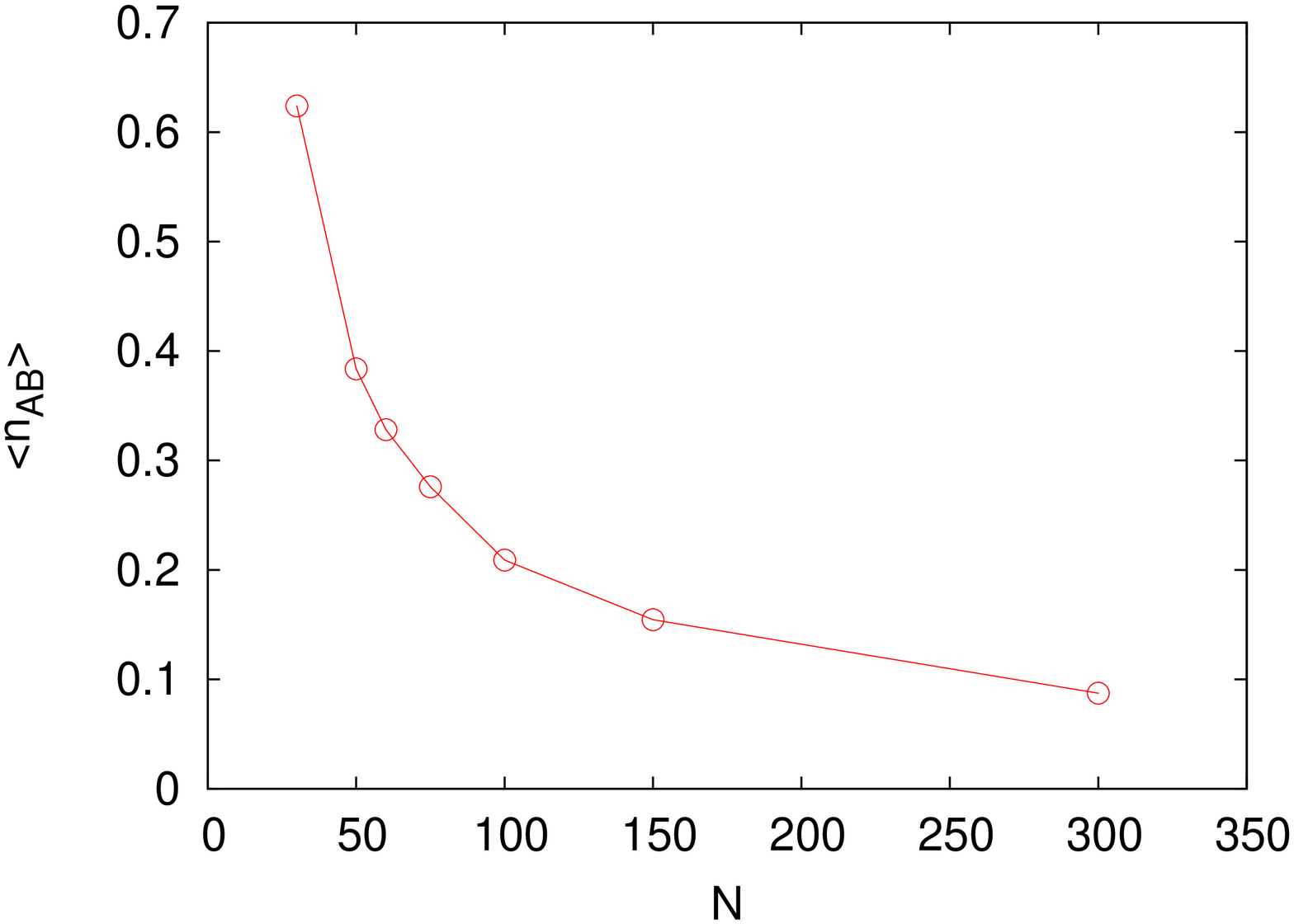}
}}}
\subfloat[][]{
\rotatebox{270}{\resizebox{!}{0.50\columnwidth}{%
  \includegraphics{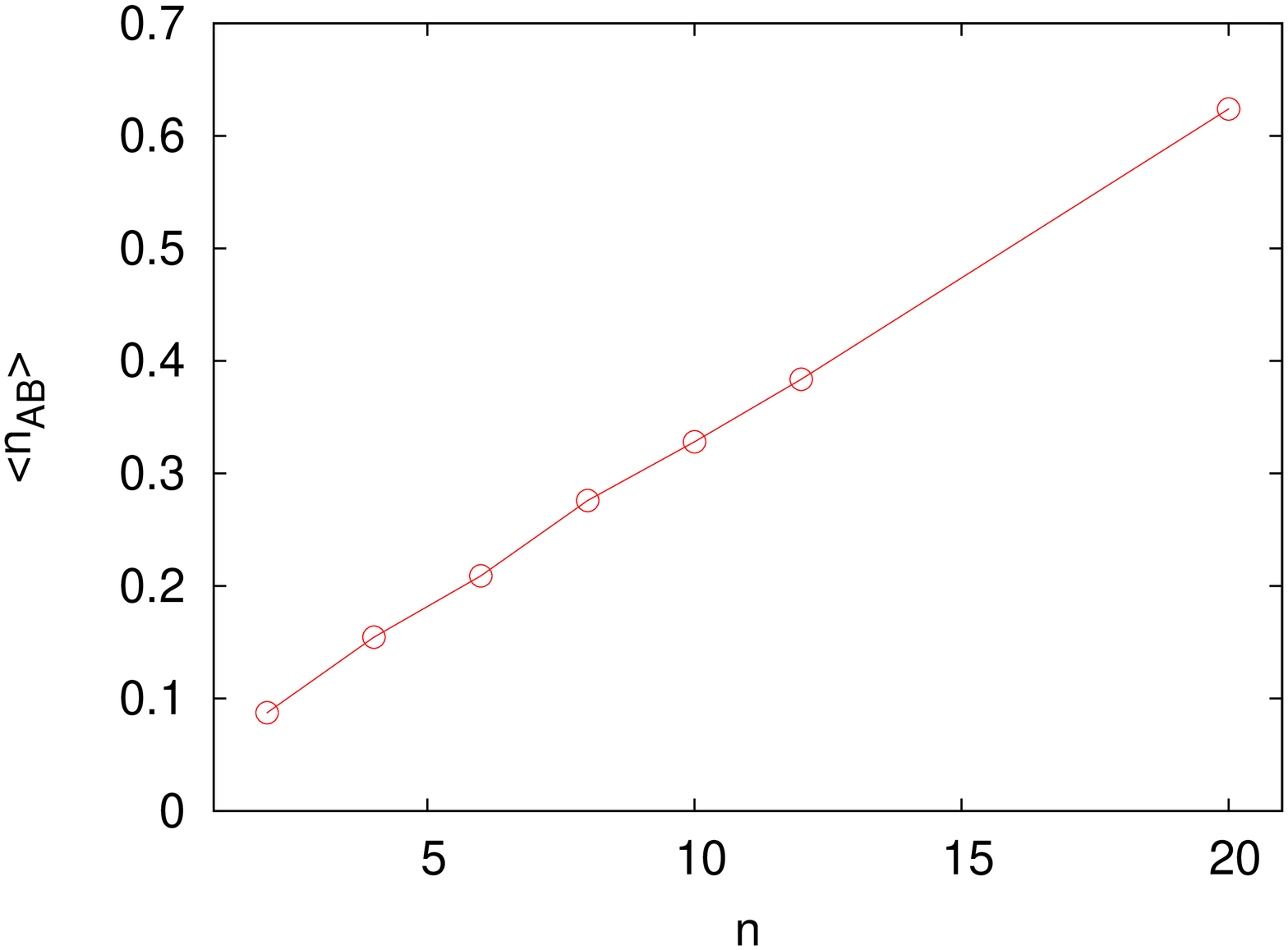}
}}}
\end{center}
\caption{\label{fig4}(Colour online)
Average number of contacts $\langle n_{AB} \rangle$ between different
type of monomers plotted versus block length
$N$ (a) or versus number of blocks $n$ (b). In all cases, $nN=600$. The range of
$N$ (or $n$) corresponds to the case, where all blocks of type A are always joined
together forming a cluster, while monomers of type B belong to another cluster and
an A-B interface between the above two clusters is formed.
}
\end{figure*}

where $\delta(\vec{x})$ is the Dirac delta function, $\vec{r}_c$
the position of the center of mass of the whole chain and
$\vec{r}_i$ the positions of all monomers irrespective of their
type (A or B). The angle brackets denote an average over all
configurations as usual. At low temperatures (e.g., in
figure~\ref{fig3}(a) $T=1.8$) the differences are more pronounced
than they appear at higher temperatures, i.e., $T=3.0$
(figure~\ref{fig3}(b)). However, the differences between a
homopolymer chain of the same length ($N=600$) and multiblock
chains ($nN=600$) at both temperatures are rather high even at
temperatures close to $\Theta$ showing that the case of multiblock
copolymers strongly deviates from the case of a homopolymer chain.
Multiblock copolymer chains are rather overally swollen in the
radial directions due to the unfavorable interactions between
neighboring A and B blocks at ``high'' temperature and all
different blocks at lower temperatures. At $T=1.8$ and for small
$n$ (or equivalently high $N$, since $nN$ is fixed in this case)
the multiblock copolymers have pronounced differences in their
densities close to the center of the macromolecule. This region of
$n$ $(2<n<20)$ for this total chain length corresponds to the case
(f) of figure~\ref{fig2}, where full phase separation of the
blocks of different type has taken place. Even though the chains
exhibit the same phase behaviour, rather pronounced differences
are observed in density profiles, showing that the interface
between A and B monomers is not the same in all cases. As $n$
increases, one would expect that the contacts between A and B
monomers at the interface should also increase in the collapse
state ($T=1.5$), and the cross-section should rather change from a
double-cylinder-like to a dumbbell-like shape~\cite{70}.
Therefore, it is relevant that we count the contacts between A and
B beads, which is expressed by the following formula~\cite{71},
\begin{equation}
\label{eq92}n_{AB} = 4 \pi\int\limits_{0}^{r_{n}}
{g}_{AB} (\Delta r) (\Delta r)^{2}d(\Delta r)
\end{equation}
where $\Delta r$ is the absolute value of the distance between two
sites of monomers $\vec{r}_{i},\vec{r}_{j}$ in the multiblock
copolymer chain, and $g_{AB}$ the corresponding radial
distribution function. Eq.~\ref{eq92} means that a pair of
monomers $(A, B)$ is defined to have a pairwise contact if their
distance is less than $r_{n}$. Similar quantities have been also
used to characterize the incompatibility between linear and star
chains in polymer melts and blends~\cite{78,79,80}. As it can be
seen from figure~\ref{fig4}, the number of contacts A-B averaged
over all conformations increases as the number of blocks $n$
increases. Also, we could see from figure~\ref{fig4}(b) that the
average number of blocks has a linear dependence on the number of
blocks $n$, although we have always the formation of only two
phase separated clusters, and the number of contacts per monomer
remains less than $1.0$. One would expect that $\langle n_{AB}
\rangle < 1.0$ could be an indication for full phase separation.
Lower values of unfavorable contacts is an indication for the
occurrence of only two phase separated clusters of A and B blocks
(full phase separation), present at this low temperature. For high
values of $n$ (figure~\ref{fig3}(a)) there are rather small
differences in the density profiles. In this regime a variation in
the number of clusters has been observed. At $T=3.0$
(figure~\ref{fig3}(b)) the differences, which were present for
small $n$ at low temperatures, have almost disappeared. Already at
$T=2.4$ the behaviour shown in figure~\ref{fig3}(b) preempts that
of figure~\ref{fig3}(a), and is an indication that the chain
leaves the collapsed state. For $T>2.4$ differences should also
result from the deviation of the overall formation from a
completely spherically formation. We postpone such detailed
discussion on these phenomena to a later communication.

\begin{figure}
\subfloat[][]{
\rotatebox{270}{\resizebox{!}{0.50\columnwidth}{%
  \includegraphics{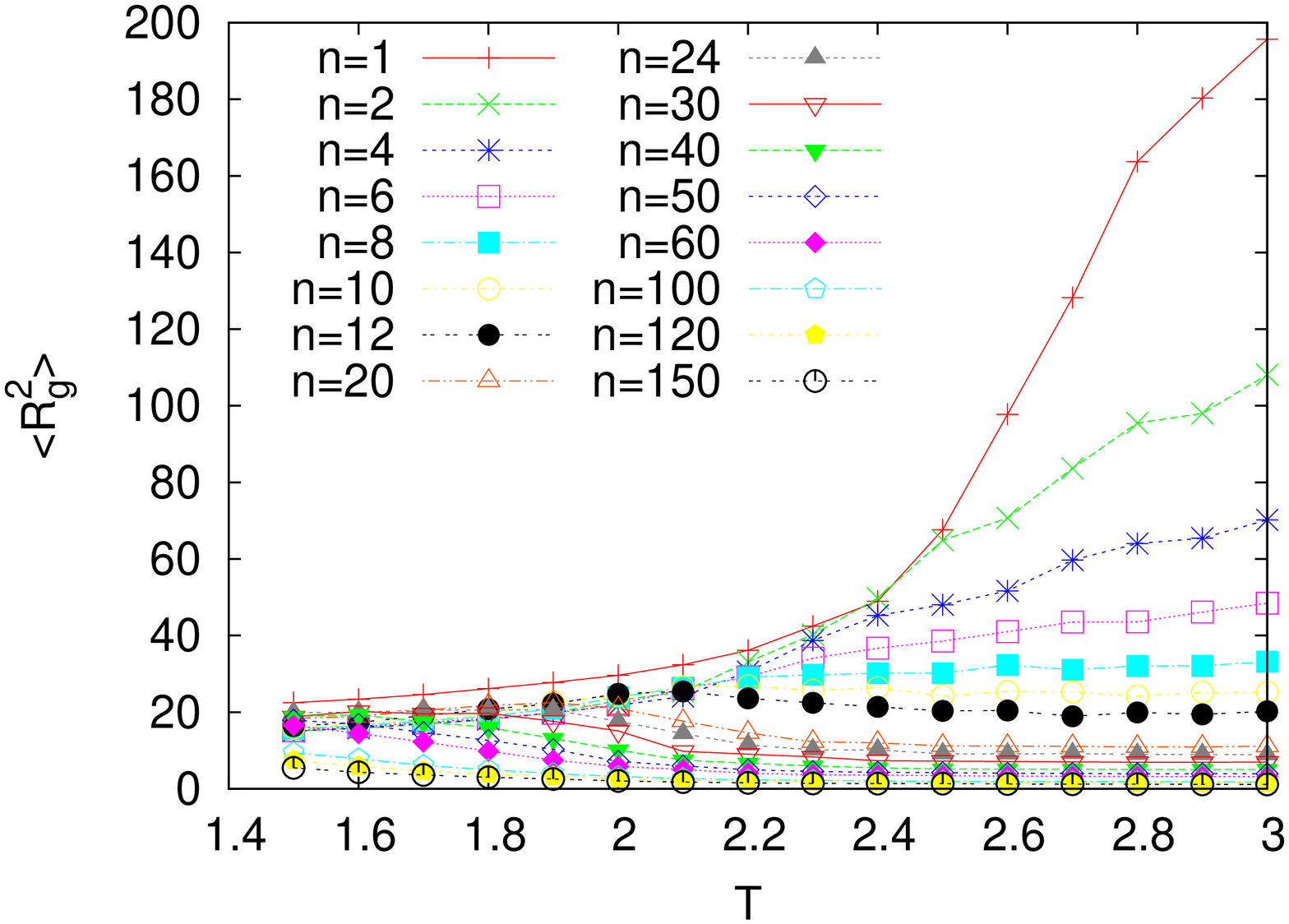}
}}}
\subfloat[][]{
\rotatebox{270}{\resizebox{!}{0.50\columnwidth}{%
  \includegraphics{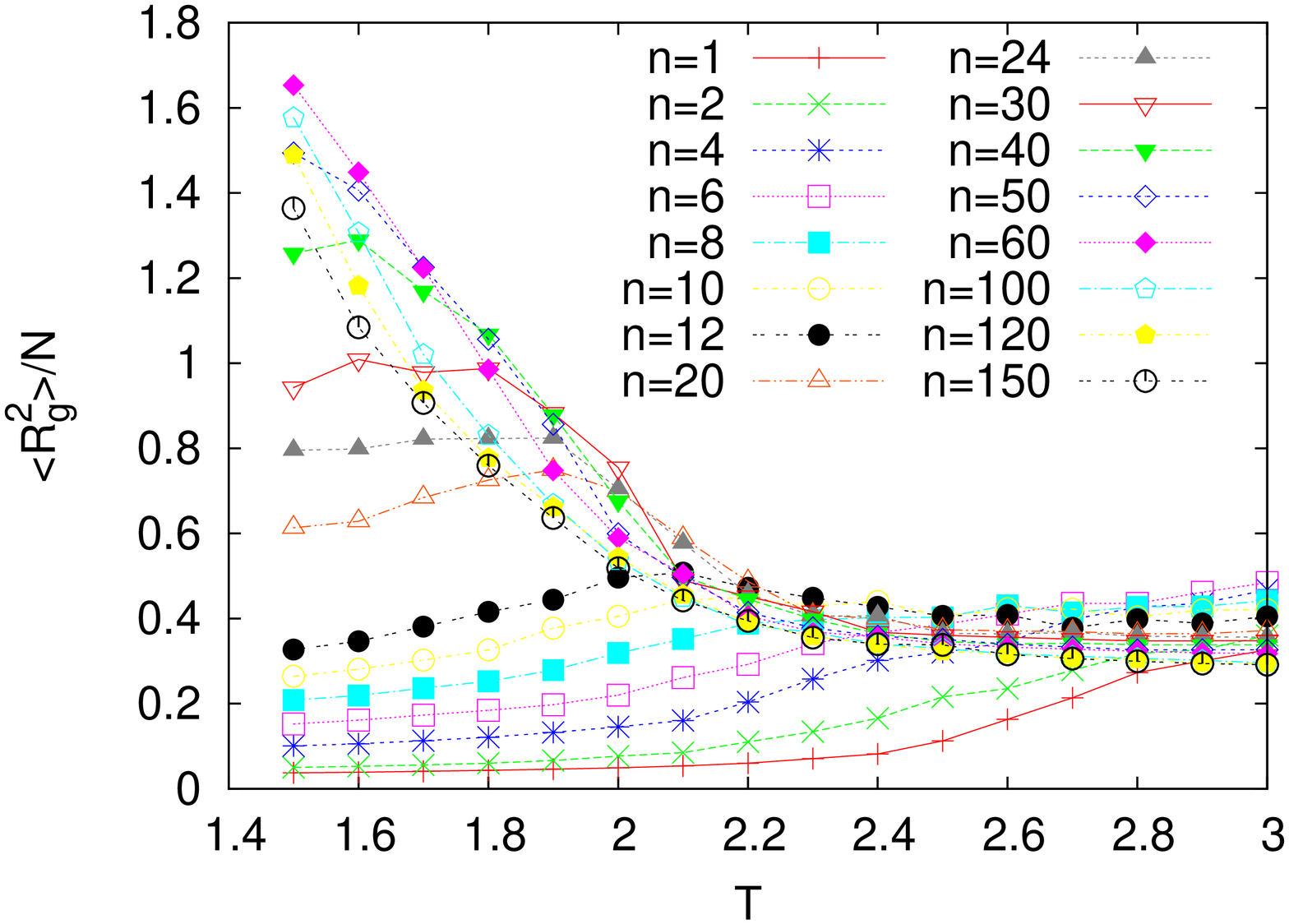}
}}}\\
\caption{\label{fig5} (Colour online) The mean square gyration
radius $\langle R^{2}_{g} \rangle$ of clusters A or B versus
temperature is shown (a). In case (b) the same data are plotted,
normalized with the block length $N$. Data are shown for different
number of blocks $n$. In all cases $nN=600$.}
\end{figure}

We turn now to the properties characterizing the size of the
clusters. Assuming that there should not be any preferential
orientation, it is natural to characterize the size of our
clusters by calculating the gyration radii of the formed clusters.
Figure~\ref{fig5} shows results for $\langle R^{2}_{g} \rangle$
and the temperature dependence of their size. For these results
the total length of the multiblock copolymer chains is $nN=600$.
At first, we can clearly see (figure~\ref{fig5}(a)) the effect of
the temperature on the homopolymer chain. A major change in the
slope of the cluster size with the temperature occurs at
temperatures close to $T=2.4$ and it is completed at temperatures
close to $T\approx 2.2$. A smooth crossover from coil-like
structures to collapsed-chain structures takes place for this
range of temperatures. This crossover becomes smoother in the
multiblock copolymer cases. Also, it seems to be slightly shifted
to lower temperatures, i.e., close to $T=2.0-2.2$. For the
multiblock copolymer chains, we can distinguish two different
behaviours. For small $n$ (these cases correspond to phase
behaviour like the one shown in figure~\ref{fig2}(f)), the size of
the clusters shows a monotonic behaviour with the temperature
variation, i.e., the size of the cluster increases monotonically
with the increase of the temperature. As $n$ increases the
variation with the temperature is smaller. Of course at high
temperatures, where rather individual blocks compose each cluster,
the size of these clusters has a straightforward dependence on the
size of the individual blocks. Moreover, the length is rather high
($n$ low, since $nN=const$ in this case) and the probability of
two clusters to join is ``high'', due to the high length of the
blocks and the high flexibility of the chain, which has a lower
number of unfavorable A-B contacts. Then, we can distinguish a
second behaviour, where the size of the clusters at higher
temperatures are smaller than the clusters at low temperatures. In
this case and at temperatures close to $\Theta$, the clusters,
which are composed of rather individual blocks, have smaller
dimensions compared to their size at low temperatures, where
blocks join together leading to the formation of clusters of
monomers of the same type. Overall, as the number of blocks $n$
increases, the behaviour of multiblock copolymers deviates further
from the homopolymer case. Moreover, the decrease of the block
length $N$ results in clusters of smaller dimensions in all cases
and at all temperatures. Only at low temperatures and in the
regime where full phase separation occurs (small values of $n$, in
the range of $N$ and $T$ presented here) we can see that the
clusters have the same size. This is the case that the multiblock
copolymer chain has only two clusters with monomers of different
type, but each cluster contains the same number of monomers A or B
for different $n$. This is the proof that the interface between A
and B clusters is mainly held responsible for the differences
arising in the density profiles for small $n$ at low temperatures.
It is natural that the size of the single cluster of the
homopolymer case should be the highest at ``high'' temperatures
(the block is the longest, $N=600$ in this case). However,
plotting our data in a different way figure~\ref{fig5}(b), we can
see that the dimensions of the clusters are on average (normalized
by $N$) higher as the dimension of the individual blocks at higher
temperatures. In such a plot one could still distinguish between
the different regimes discussed above.

\begin{figure*}
\begin{center}
\subfloat[][]{
\rotatebox{270}{\resizebox{!}{0.50\columnwidth}{%
  \includegraphics{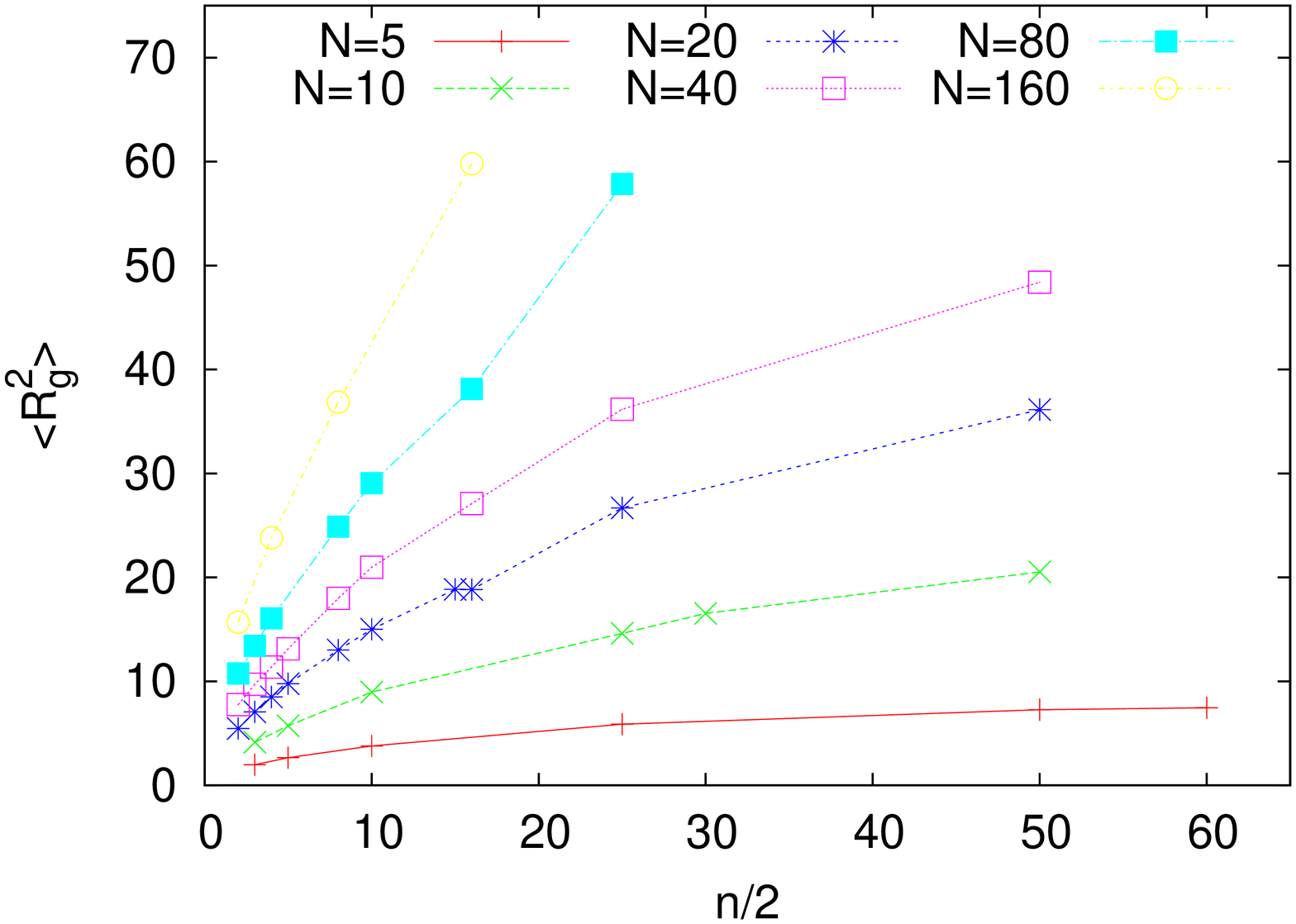}
}}}
\subfloat[][]{
\rotatebox{270}{\resizebox{!}{0.50\columnwidth}{%
  \includegraphics{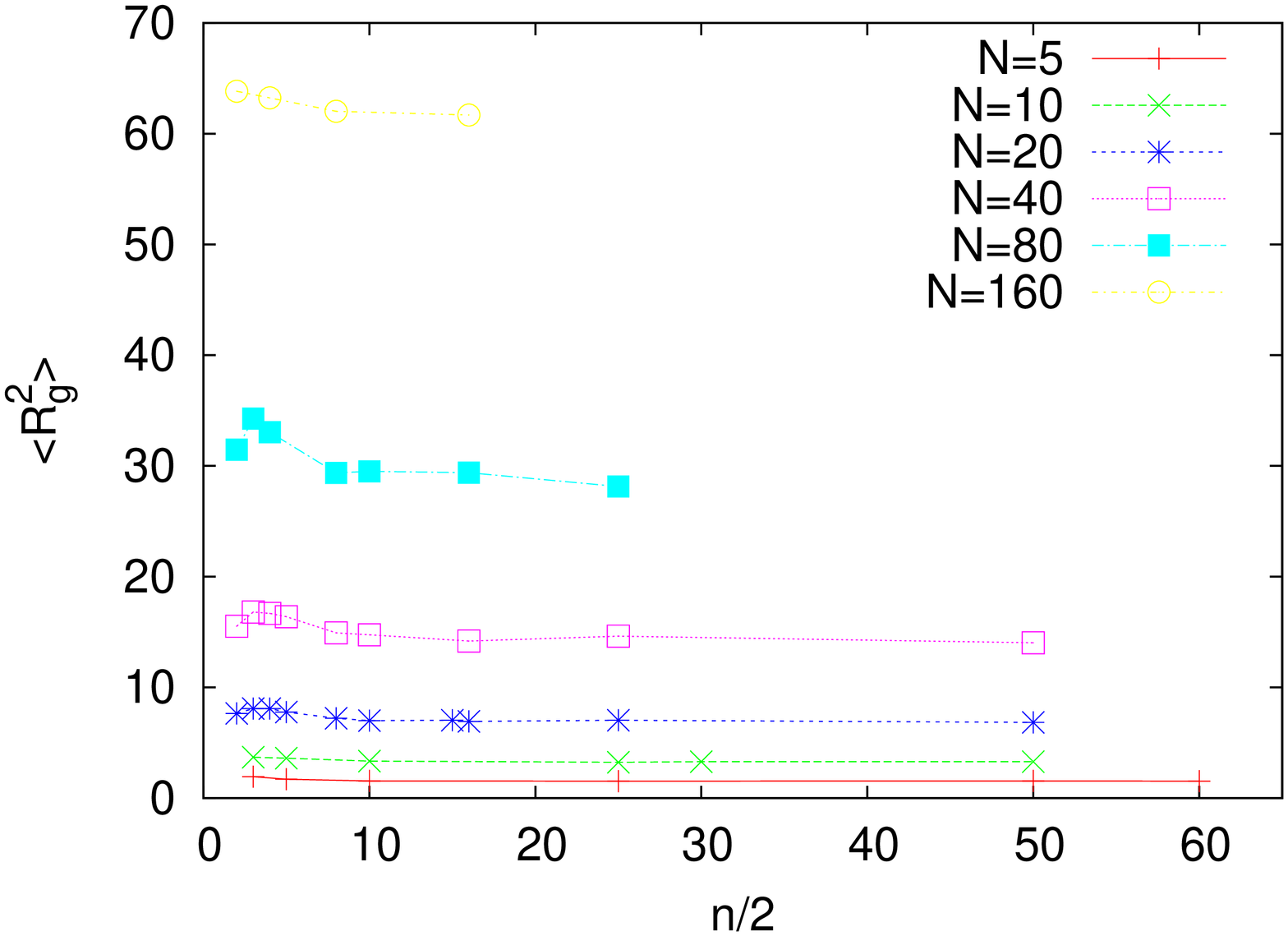}
}}}
\end{center}
\caption{\label{fig6} (Colour online) The mean square gyration
radius $\langle R^{2}_{g} \rangle$ of clusters A or B plotted
versus the number of A or B blocks ($n_{A}=n_{B}=n/2$) for different block
lengths. Two different temperatures are shown: $T=1.5$ (a) and
and $T=2.7$ (b).}
\end{figure*}

In figure~\ref{fig6} we present results for the dependence of the
average size of the cluster formations on the number of blocks
$n/2$ of A-type monomers. Of course, as it has been already
discussed the results are the same for B-type monomers due to the
symmetry of our model. Therefore, keeping constant the block
length $N$, we altered the number of blocks $n$.
Figure~\ref{fig6}(a) presents such results at $T=1.5$. When $N$ is
small (e.g., $N=5$), a rather small variation with the number
$n/2$ of A or B blocks is seen. As the block length increases the
slope of the curves also increases, showing that the variation
with the number of blocks $n$ is high. It should be conceivable
that also for high $N$ ($N=160$) a plateau-like regime could be
reached as in the case of $N=5$, such that the system departs from
the full phase separation regime. Yet, we are not able to access
this regime with our simulations. By increasing the temperature,
for instance $T=2.1$ (not shown here), the cases of small $N$ do
not show any dependence with the variation of the number of
blocks, whereas for high $N$ the linear dependence regime becomes
smaller. We know that phase separation is favored from higher
block lengths. Therefore, blocks of small $N$ start to behave
independently (leave a cluster) already at a lower temperature. At
temperatures close to the $\Theta$ temperature (above $T=2.4$,
i.e., $T=2.7$, figure~\ref{fig6}(b)) cluster formations containing
two or more blocks of the same type of monomers are hardly formed.
Therefore, one should hardly observe any dependence on the number
of blocks $n$ and the size of the clusters should rather reflect
the size of individual blocks However, we remind the reader that,
even at these high temperatures, blocks with monomers of the same
type do occasionally form clusters, and therefore the size of
these clusters should not be considered absolutely as the size of
the individual blocks.

\begin{figure}
\subfloat[][]{
\rotatebox{270}{\resizebox{!}{0.50\columnwidth}{%
  \includegraphics{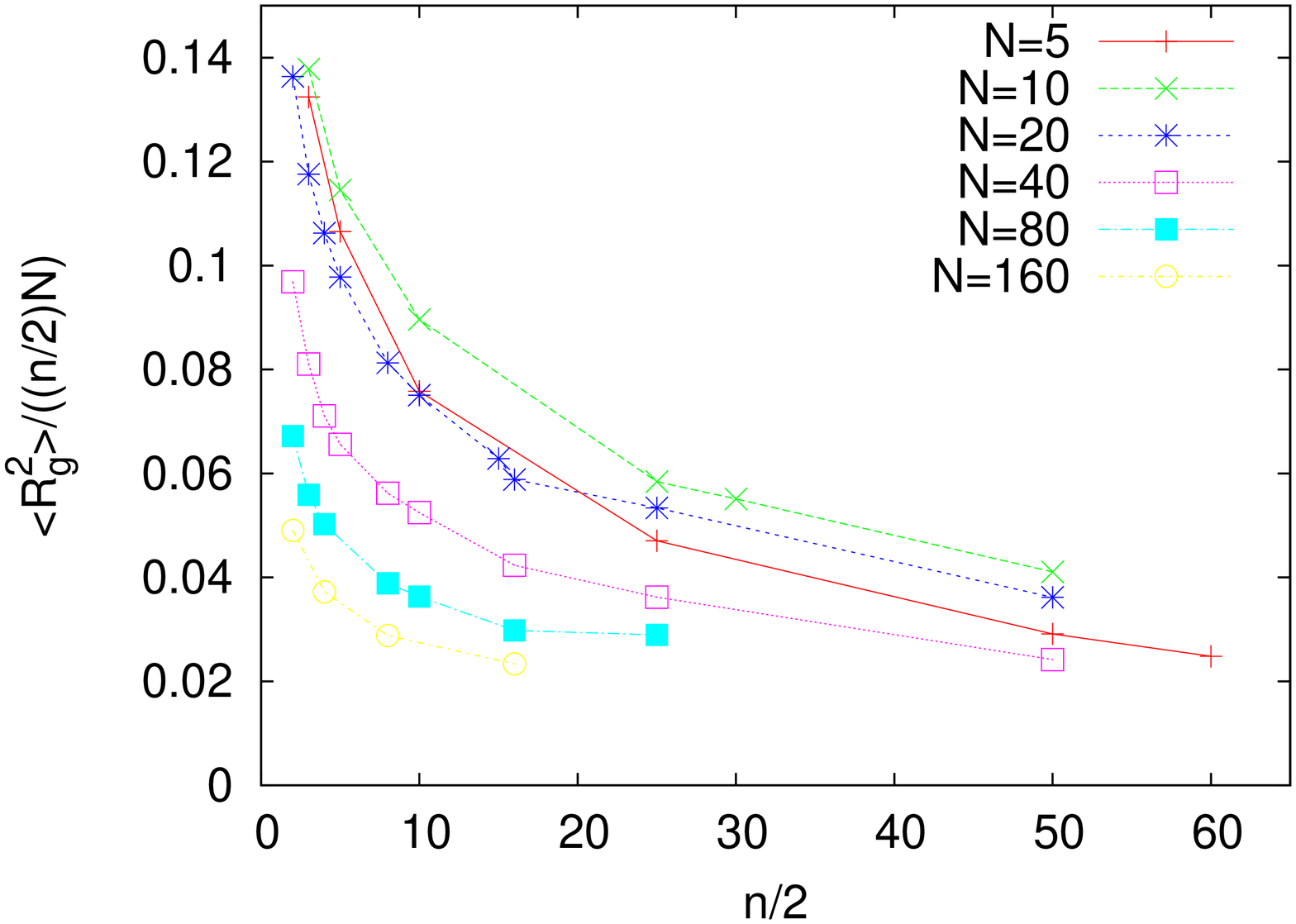}
}}}
\subfloat[][]{
\rotatebox{270}{\resizebox{!}{0.50\columnwidth}{%
  \includegraphics{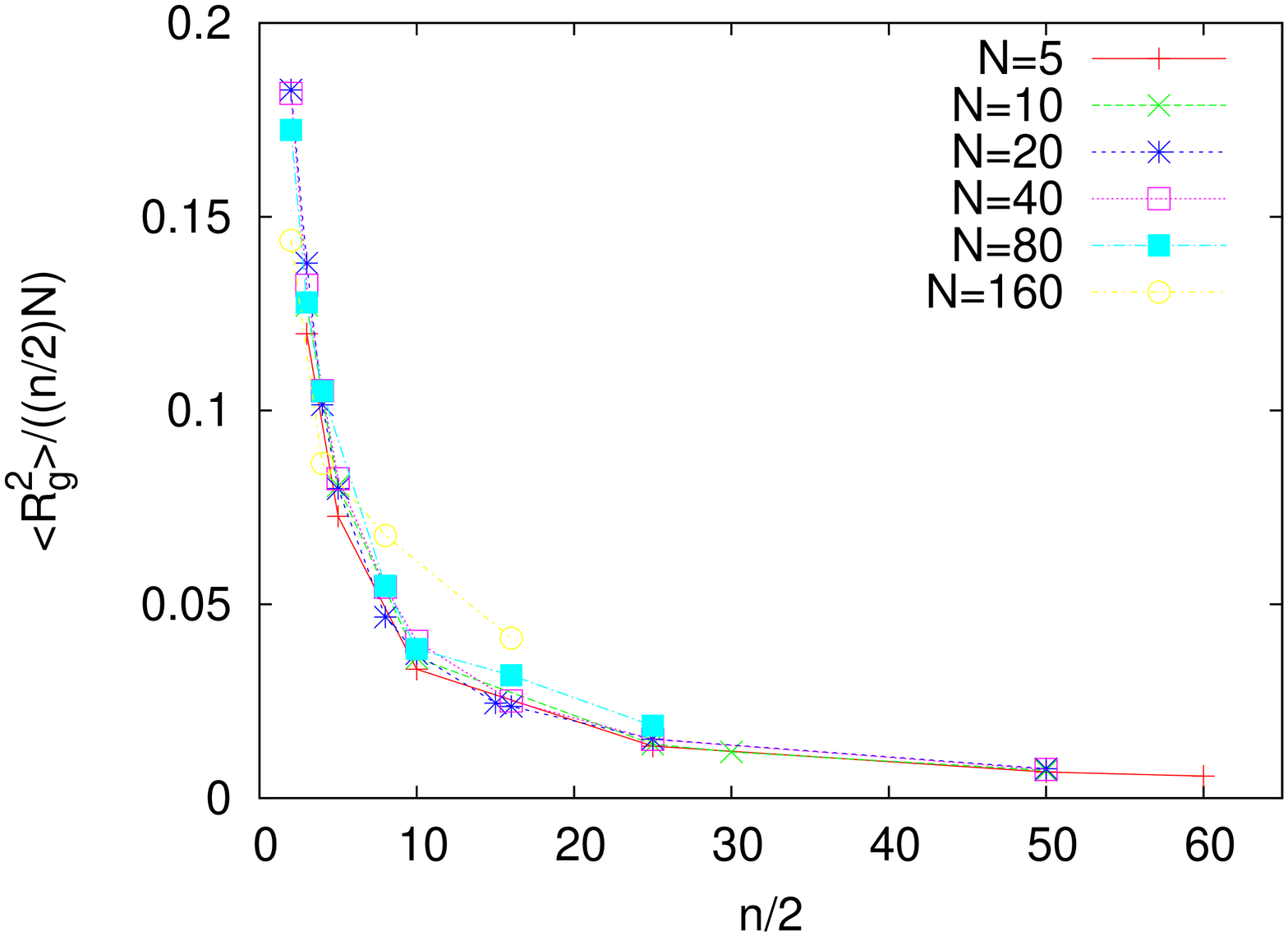}
}}}
\caption{\label{fig7} (Colour online) The mean square gyration
radius $\langle R^{2}_{g} \rangle$ of clusters A or B divided by
the number of A or B blocks ($n_{A}=n_{B}=n/2$) and the block
length ($N$) plotted versus this number of A or B blocks for
different block lengths. Two Different temperatures are shown: $T=1.5$
(a) and $T=2.4$ (b).}
\end{figure}

It would be desirable to plot our results in such a way that the
regime of rather individual clusters could be distinguished from
the regime of collapsed chain. Such plots are shown in
figure~\ref{fig7}. At low temperatures (i.e., $T=1.5$,
figure~\ref{fig7}(a)) and for block lengths $N$ below $20$ ( for
these systems a variation in the number of clusters is observed
and occasionally two only separated clusters of different monomers
are observed~\cite{50,51}) the curves do not show any systematic
behaviour. It is only for higher block length $N$, where the block
length high enough to lead to phase separation between different
blocks, that we can clearly see the different behaviour. A
``plateau regime'' is accessed faster for higher $N$. As the
temperature increases (above $T \approx 2.4$), we reach a
temperature where the curves collapse onto a single curve for the
different block lengths $N$ showing the same behaviour with the
number of A or B blocks $n/2$. Our claim for a universal
temperature boundary around $T=2.4$ (figure~\ref{fig7}(b)) seems
to be correct. The effect of the solvent quality is proven to play
an important role below this temperature, for all the multiblock
chains. Further increase of the temperature increases the overall
size of the chains. Thus, at $T=3.0$ (not shown here) the curves
are still on top of each other as in figure~\ref{fig7}(b), but
only a small change in the slope of the curves with the number of
blocks $n/2$ can be observed. This shows that the increase of the
temperature has the effect of weakening further the dependence on
the number of blocks $n$. Also, increase of the temperature shifts
the curvatures slightly to higher values of $(\langle R^{2}_{g}
\rangle)/(Nn_{A})$, as it is expected.

\begin{figure*}
\begin{center}
\subfloat[][]{
\rotatebox{270}{\resizebox{!}{0.50\columnwidth}{%
  \includegraphics{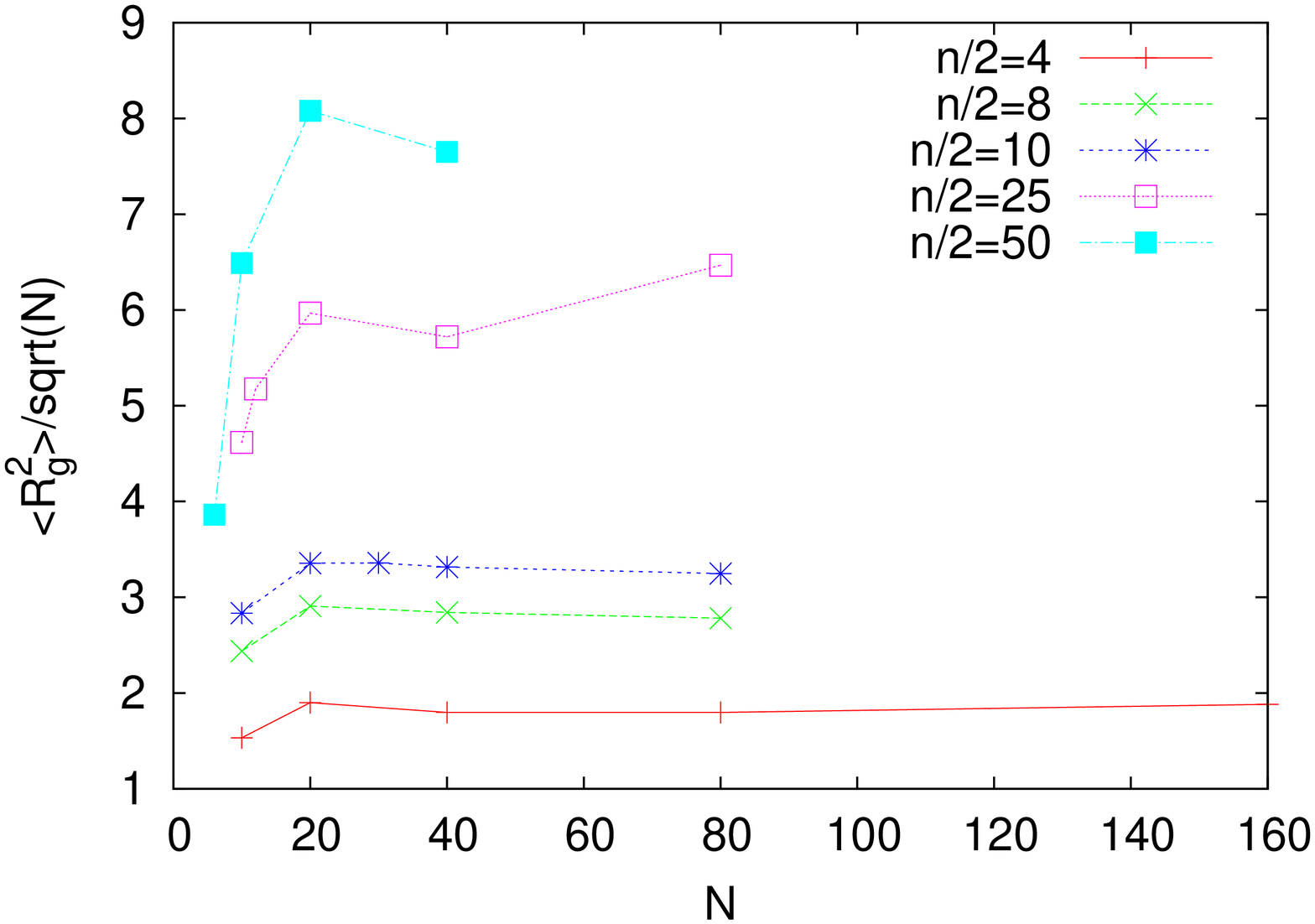}
}}}
\subfloat[][]{
\rotatebox{270}{\resizebox{!}{0.50\columnwidth}{%
  \includegraphics{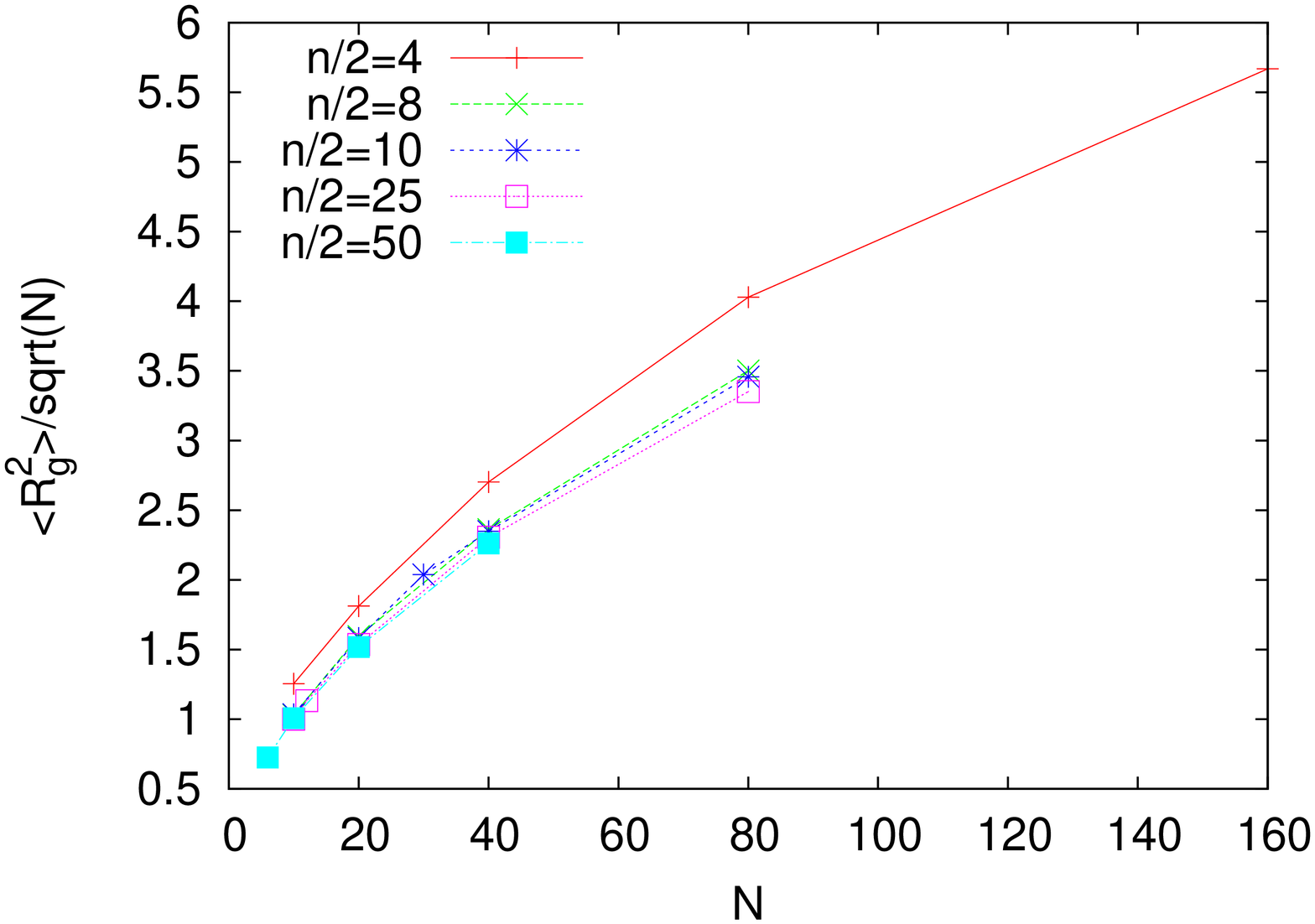}
}}}
\end{center}
\caption{\label{fig8} (Colour online) The mean square gyration
radius $\langle R^{2}_{g} \rangle$ of clusters A or B divided by
the square root of the block length ($N$) plotted versus block
length for different number of blocks as indicated. Two different
temperatures are shown: $T=1.5$ (a) and $T=3.0$ (b).
}
\end{figure*}

We can present correspondingly the dependence of the cluster sizes
on the block length for different multiblock copolymer chains
composed of different number of blocks ($n_{A}=n_{B}=n/2$). Such
plots are presented in figure~\ref{fig8} for two different
temperatures, i.e., a low temperature where the chain is fully
collapsed, and a temperature close to the $\Theta$ temperature. We
have seen that full phase separation between blocks can not take
place when the block length is lower than $N=20$ in the range of
temperatures which are studied here. In the regime of full phase
separation as it has been discussed in this study (at rather low
temperatures) the size of the clusters does not depend on the
block length $N$ for different number of blocks, when the
normalization of figure ~\ref{fig8} is chosen. Below $N=20$, full
phase separation cannot take place, a variation in the number of
clusters is observed, and higher dependence is seen on the block
length $N$ for higher number of clusters, until a plateau is
reached for $N>20$. This behaviour is rather universal for
multiblock chains of different number of blocks. Then, at $T=3.0$
our data show the same behaviour exhibiting an almost linear
dependence on $N$ for all the cases of different $n$. However, for
small number of blocks length ($n/2=4$) the curve is shifted. For
higher values of blocks $n$ this shift is smaller as this number
of blocks increases. This behaviour is compatible with the picture
obtained from overall properties, as it is, for example, the
density profiles of figure~\ref{fig3}(b).

\begin{figure*}
\begin{center}
\subfloat[][]{
\rotatebox{270}{\resizebox{!}{0.50\columnwidth}{%
  \includegraphics{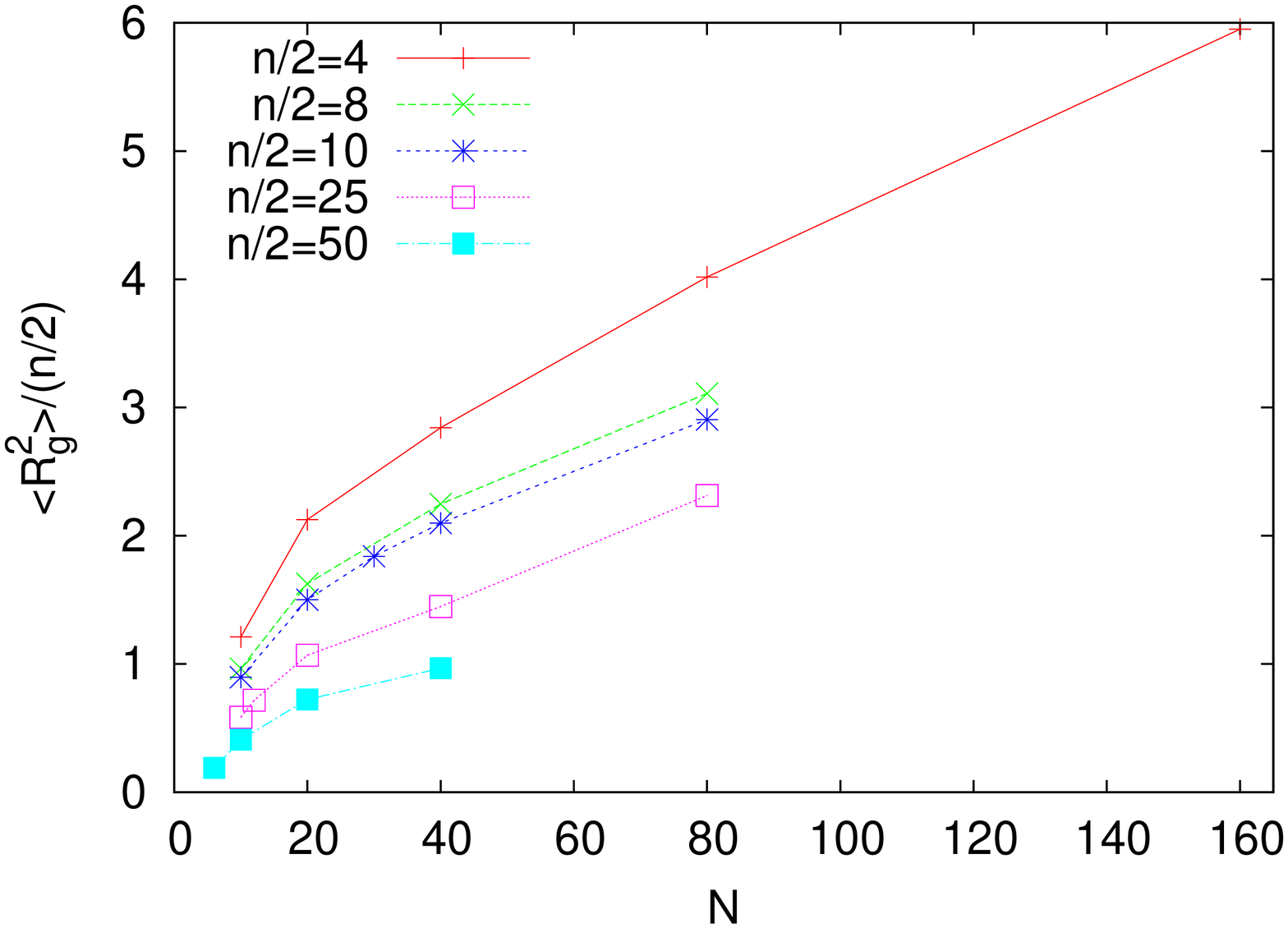}
}}}
\subfloat[][]{
\rotatebox{270}{\resizebox{!}{0.50\columnwidth}{%
  \includegraphics{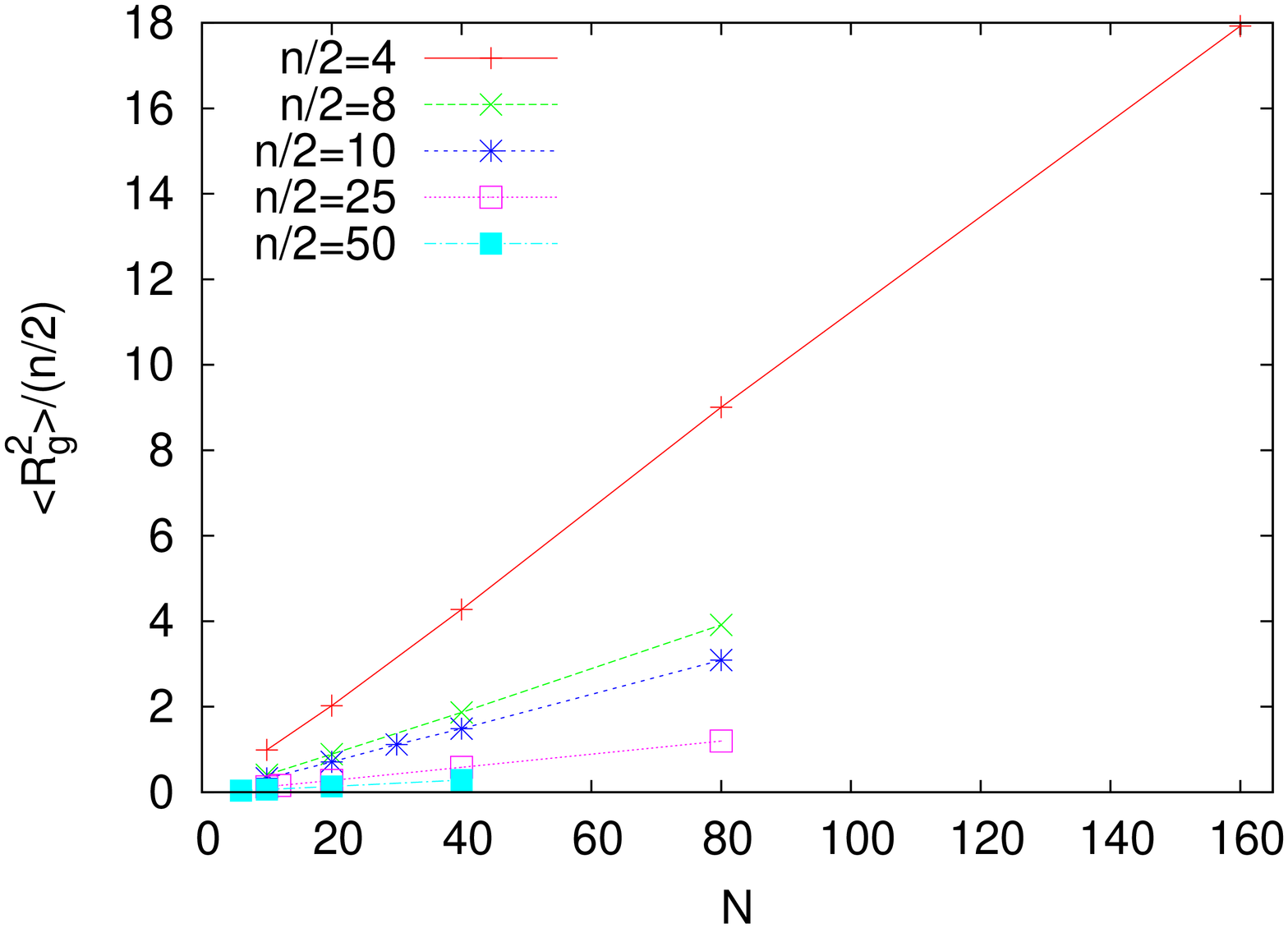}
}}}
\end{center}
\caption{\label{fig9} (Colour online) The mean square gyration
radius $\langle R^{2}_{g} \rangle$ of clusters A or B divided by
the number of blocks A or B ($n/2$) plotted versus block length
for different number of blocks as indicated. Two different
temperatures are shown: $T=1.5$ (a) and $T=3.0$ (b).
}
\end{figure*}

It is interesting to augment our discussion by dividing our data
by the number of blocks ($n_{A}=n_{B}=n/2$, figure~\ref{fig9}). At
lower temperatures (i.e., $T=1.5$) a universal boundary between
two linear regimes is seen; when $N$ is small i.e., $N<20$. Then a
different regime is reached, where a linear dependence with a
different slope from the first linear regime is seen. As the
temperature increases close to $\Theta$, where the clusters are
mainly composed by individual blocks, we can clearly see a single
linear behaviour extending over all the considered block lengths
$N$. When the number of blocks $n$ is small, the slope describing
the dependence on the block length $N$ is higher. In this case
blocks of monomers of the same type have higher probability to
cluster occasionally. As the number of blocks $n$ increases, this
probability of occasional clustering globally decreases as the
number of unfavorable interactions decrease. Such a linear
behaviour shows that that the blocks individually behave
``locally'' like homopolymer chains, and it is only the A-B
contacts between connected blocks that control such behaviour.
Plots like figure~\ref{fig9} provide an indication of the
proximity of the $\Theta$ temperature for the chains.

\section{Concluding remarks}
\label{conclusions}

In this study, we have investigated the static properties of a
single multiblock copolymer chain under poor solvent conditions.
The interactions were chosen symmetrically, as well as the
structural parameters of the chains, i.e., the length of the blocks
with monomers of type A or B were equal, and the number of blocks
A was equal to the number of blocks B, while the blocks of
different type alternated along the chain. We used standard
molecular dynamics simulations of a bead-spring model for our simulations, and
we focused our discussion on the dependence of the size of the
cluster formations occurring at low temperatures on the varied
parameters, i.e., the block length $N$, the
number of blocks $n$, and the temperature $T$ which was used to
tune the quality of the solvent. Our analysis was presented in the
context of recent results on the phase behaviour of such
macromolecules. We showed that the number of contacts at the
interface between A and B monomers, in the case where the
multiblock copolymer chain at a low temperature (e.g., $T=1.5$) is
always split in two microphase separated clusters composed of
blocks of the same type of monomers, varies linearly with the
number of blocks and it is lower than $1.0$ ($nN$ was kept
constant).

Accordingly, we studied the dependence of the clusters' size on
the temperature discussing our results in the context of recent
results on the phase behaviour of such macromolecules. We could
clearly distinguish the different regimes relating to the
different phase behaviours. The dimensions of the clusters as a
function of the number of blocks $n$ and the block length $N$ were
independently studied. A comprehensive discussion of the
dependence of the clusters' size on the $n$ was given, discussing
also the effect of the temperature. Furthermore, we showed
(figure~\ref{fig7}) that our data collapse onto a single curve
when the chains leave the collapsed state and gradually adopt
coil-like conformations. Thus, we could use such plots to identify
this boundary. This occurs at temperature $T \approx 2.4$. Then,
the dependence of the size of the formed clusters on the block
length $N$ was presented. We showed that at lower temperature the
collapsed state results in two different linear regimes with a
smooth crossover between them. At temperatures close to $\Theta$
temperature a linear dependence on the block length $N$ is seen
for all block lengths with a different slope, i.e., as the number
of blocks increases, there is a smaller dependence on the block
length $N$, when the data for gyration radius are normalized with
the number of blocks. Our results, in combination with recent
results on the phase behaviour of symmetric linear multiblock
copolymers, provide a complete picture of the behaviour of the
cluster formations in such macromolecules for the range of
parameters accessible to simulations.

\end{document}